# Product Evaluation In Elliptical Helical Pipe Bending

Wasantha Samarathunga[1], Masatoshi Seki[2], Hidenobu Saito[3], Ken Ichiryu[4], Yasuhiro Ohyama[5]

*[1,5]Tokyo University of Technology*
*Tokyo, Japan.*

*[2,3,4]Mechatronics Research Laboratory*
*Kikuchi Seisakusho Co. Ltd.*

*Abstract*— This research proposes a computation approach to address the evaluation of end product machining accuracy in elliptical surfaced helical pipe bending using 6dof parallel manipulator as a pipe bender. The target end product is wearable metal muscle supporters used in build-to-order welfare product manufacturing. This paper proposes a product testing model that mainly corrects the surface direction estimation errors of existing least squares ellipse fittings, followed by arc length and central angle evaluations. This post-machining modelling requires combination of reverse rotations and translations to a specific location before accuracy evaluation takes place, i.e. the reverse comparing to pre-machining product modelling. This specific location not only allows us to compute surface direction but also the amount of excessive surface twisting as a rotation angle about a specified axis, i.e. quantification of surface torsion. At first we experimented three ellipse fitting methods such as, two least-squares fitting methods with Bookstein constraint and Trace constraint, and one non-linear least squares method using Gauss-Newton algorithm. From fitting results, we found that using Trace constraint is more reliable and designed a correction filter for surface torsion observation. Finally we apply 2D total least squares line fitting method with a rectification filter for surface direction detection.

*Keywords*— 6DOF Parallel Manipulator, Elliptical Helical Bending, Least-Squares Fitting, Surface Torsion Observation, Surface Direction Detection

## I. INTRODUCTION

Designing build-to order manufacturing of wearable metal supporters is a very challenging task and a new trend in welfare industry. In research [1] we proposed a novel precision pre-machining modelling approach to achieve elliptical helical pipe bending to reduce excessive surface twisting during pipe bending process. The bending object is an elliptical shaped metal pipe (denotes elliptical pipe) as the cross section of pipe is in the shape of an ellipse.

The need of the wearable muscle supporters are been researched in neurophysiology and many rehabilitation areas. The Kozyavkin Method In [4] is famous as one of such prospective applications using wearable muscle supporters. The following figure 1 which is cited in research [1] briefly explain the main research project that based for this research, which is about building wearable metal muscle supporters using elliptical pipe bending into a semi elliptical helical shape. In this paper we address the post-machining product

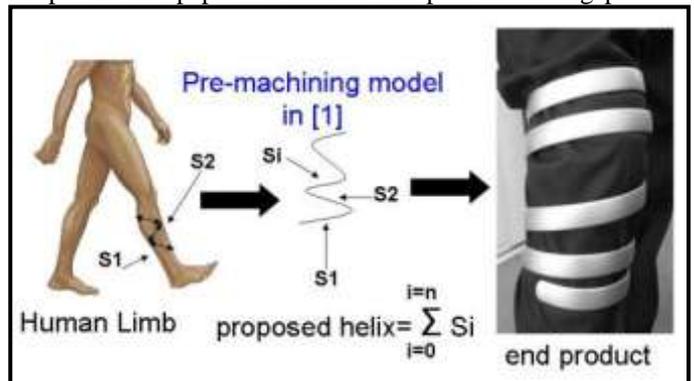

evaluation model.

Figure 1. Designing wearable metal muscle supporters

The product computation is consisting of two models, a pre-machining model [covered in 1] and a post-machining model in this paper denotes product evaluation model. Figure 2 illustrates the both models and figure 3 illustrates the usage

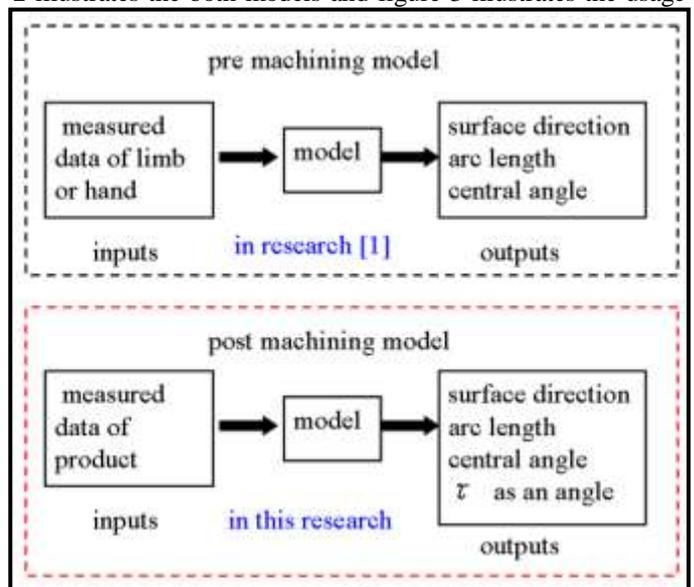

of the models.

Figure 2. Product computation in models

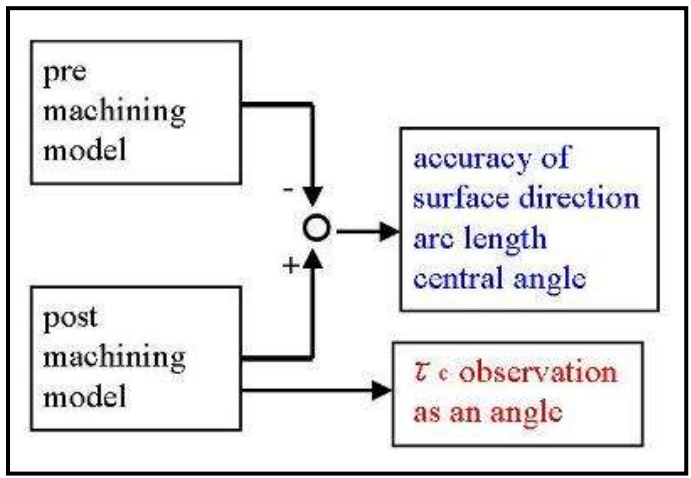

Figure 3. Usage of models

The out puts of proposed modelling are surface direction, arc length and central angle. The torsion is a key parameter during helical bending. But due to structural dependency of the bender bellow is addressed as a precision indicator. When a coordinate system placed in such a way that the Z axis (denotes $Z_w$) is placed along the centre of the final product and the helical bending can be considered as rotation about this $Z_w$ axis. X axis (denotes $X_w$) and Y axis (denotes $Y_w$) creates a plane at the bottom of the product where the bending starts.

The pipe bender used in machining is consist of rotation die mounted on a 6DOF parallel manipulator that can rotate the front end of the pipe with six degree of freedom, a fixed die to hold the pipe from the other end and a pushing mechanism of pipe at controlled speed. The posture of the end-effecter controls the assumed centre line of desired shape to be bended while the actual bending forces are applied to the surface of the pipe. The bender is illustrated in the following figure 4 which is also cited in [1].

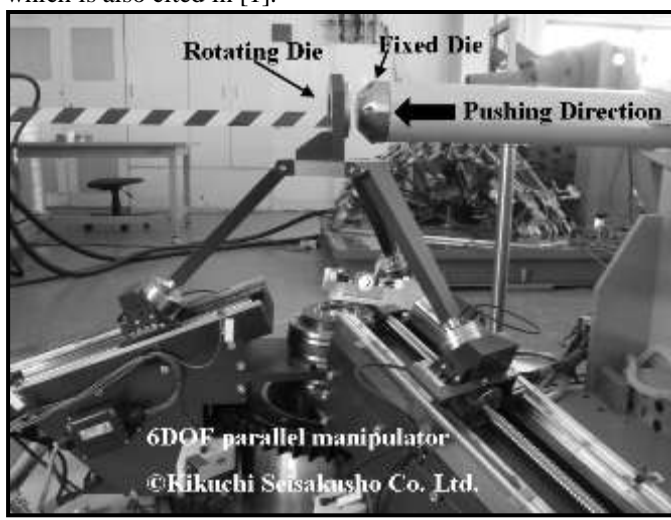

Figure 4. 6DOF pipe bender

In [1] we took a discrete pre-machining modelling to avoid excessive surface twisting during bending. Let τ represents the torsion of pipe surface and $τ_c$ represents the torsion of the pipe centre. The above mechanism only allows us to apply forces to the pipe surface rather than pipe centre by changing the posture of end-effecter. Excessive surface twisting can be represented as the cases where $τ_c ≠ τ$. In this paper we observe this τ as a precision indicator.

With the brief explanation above the research objective of this paper becomes a 3D ellipse detection problem in end product evaluation.

There are three well known methods of ellipse detection methods calls geometrical [5, 6, and 7], algebraic and Hugh Transform based [8, 9, and 10]. Researches [2 and 3] illustrate practical representations of both algebraic and geometrical methods. Since Hugh Transform methods are known to be low in computation effectiveness, we adopt geometrical and algebraic methods as in [2] and [3], compare the results of ideal data (noise free data) to select the featured fitting method. Then design a correction filter for the front end before apply actual product evaluation data with possible noise.

## II. PRODUCT EVALUATION MODEL DESIGN

The following figure 5. Illustrate the modelling strategy of the proposed approach. Our helical bending can be considered consist of sum of arcs rotated about $Z_w$. Therefore the equations of arc could be used to calculate the missing parameters. It is assumed that radii, central angle (an angle about $Z_w$) and arc length of a particular arc could be easily calculated from the means of measured surface data, and is opted here. The detection of surface direction and surface torsion τ are the featured in this computation object. We use 2D ellipse detection and 2D line fitting after applying inverse rotations and translations to a specific evaluating point.

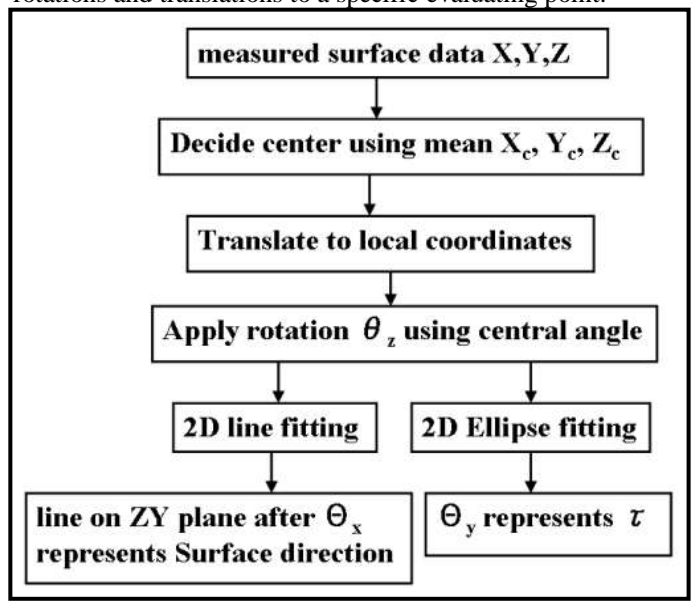

Figure 5. Proposed modelling strategy

At this specific evaluation point the target ellipse could be considered to be in the following form,

If the surface direction is zero ($\theta_x=0$), then
(1) Semi major axis of elipse is along Z axis.
(2) Semi minor axis of ellipse is along X axis.
(3) Ellipse will be in ZX plane.

If the surface direction is not zero but a rotation angle about X axis ($\theta_x$), this will result in,
(1) $-\pi/2 < \theta_x < \pi/2$
(2) Semi major axis leaves ZX plane
(3) Semi minor axis remains along X axis
(4) A rotation angle of projected line in ZY plane represents surface direction

These features are explained in figure 6. Using the above properties we can apply 2D line fitting based on orthogonal distance regression calculation as in [11] to locate surface direction as an angle $\theta_x$. This angle (denotes $\alpha$ in [1]) is the acute counterpart of helix angle.

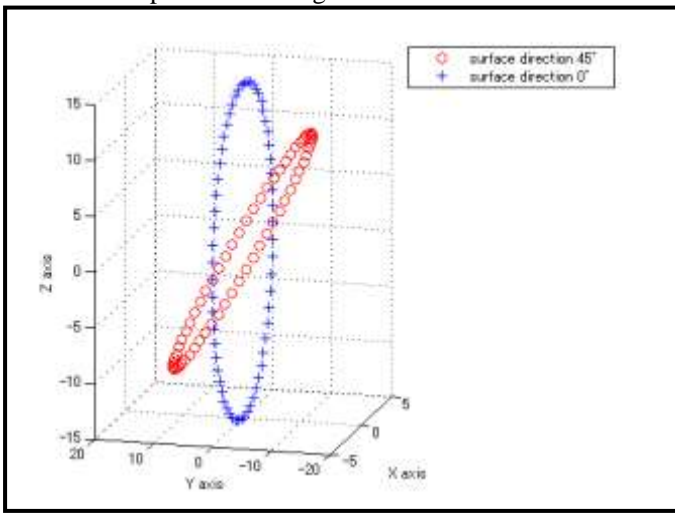

Figure 6. Surface direction as $\theta_x$

At this evaluation point surface torsion $\tau$ can be represented quantifiably as a rotation angle about Y axis, i.e. $\theta_y$. This angular representation is a relative measure and only be meaningful in continuous data evaluation of adjacent measurements. Figure 7 illustrate this rotation about Y axis.

In the following sections we address these surface torsion observation by using 2D ellipse fitting and surface direction detection using 2D line fitting respectively. Surface torsion observation starts with a comparison of ellipse fitting methods and filter design to rectify the observation values. The surface direction detection is addressed ad a 2D line fitting and angle calculation.

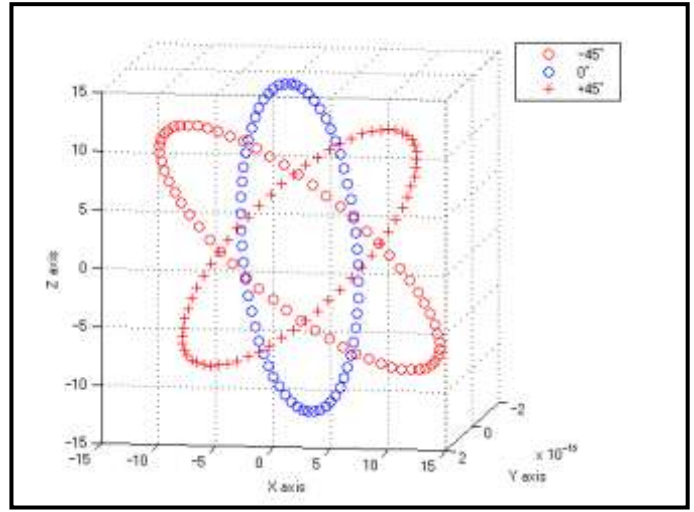

Figure 7. Change of $\tau$ as $\theta_y$

### III. SURFACE TORSION OBSERVATION

In this section of the paper is aimed to locate an ellipse fitting method with reliability. Despite of many researches and existing methods on fitting ellipses, to locate a reliable method for case like ours indeed is a difficult task than it sounds. This is due to the fact that fitting accuracy depends on constraints and different constraints act differently under different Euclidean transforms.

If the 2D equation of ellipse is represented by the following equation (1),

$$F(x) = x^T A x + b^T x + c = 0 \quad (1)$$

where A is a 2x2 matrix and x contains 2D coordinates of the relevant plane.

The least squares fitting problem could be addressed in the form of minimize the following,

$$\sum F(x)^2 \quad (2)$$

We apply two linear least squares ellipse fitting methods that minimize algebraic distances and one nonlinear least squares method that minimize the geometric distance as follows.

(1). Bookstein constraint (linear)
This apply the constraint in equation (3) bellow,

$$\lambda_1^2 + \lambda_2^2 = 1 \quad (3)$$

(2). Trace constraint (linear)
This apply the constraint in equation (4),

$$\text{Trace}(A) = \lambda_1 + \lambda_2 = 1 \quad (4)$$

(3). Gauss-Newton method (nonlinear)

The following Figures 8, 9 and 10 illustrates the fitting results of $\theta_y$ for ideal data.

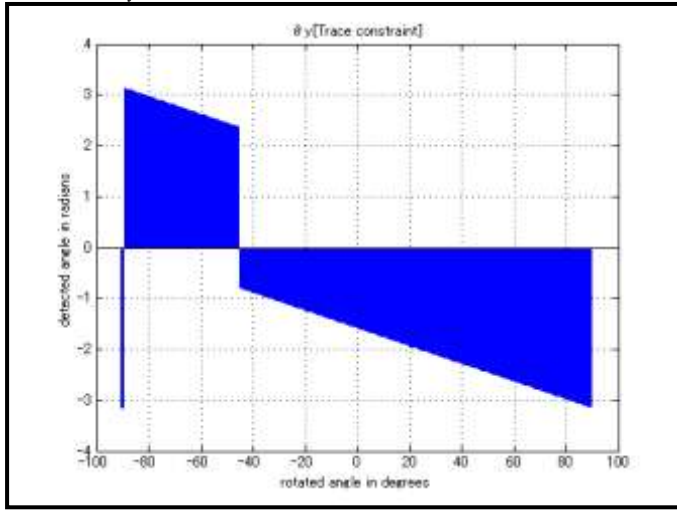

Figure 8. Detected $\theta_y$ (Trace constraint)

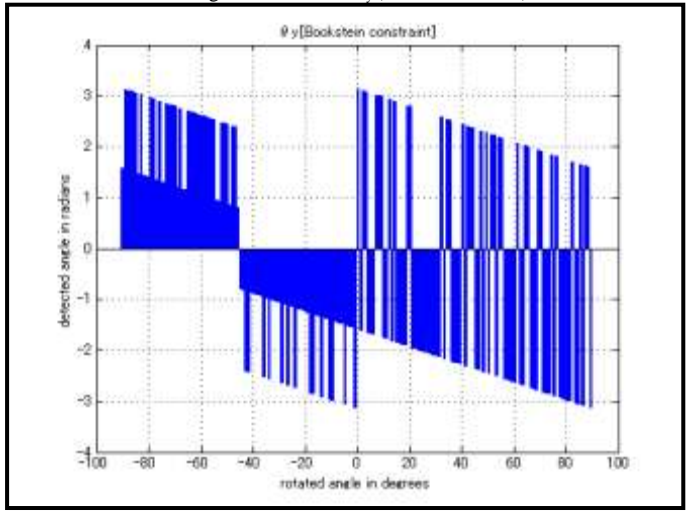

Figure 9. Detected $\theta_y$ (Bookstein constraint)

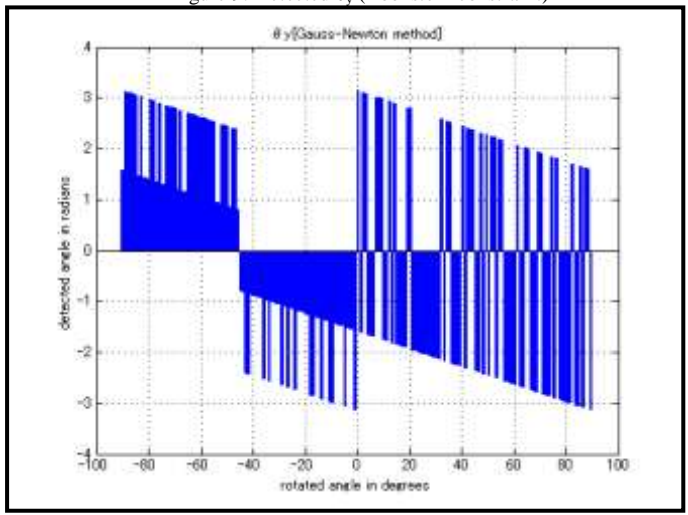

Figure 10. Detected $\theta_y$ (Gauss-Newton method)

The fitting results show that 'Trace constraint' which is illustrated in figure 8 is more stable to proceed with than the other two methods. Hence we design a discrete filter to rectify the output of 'Trace constraint' for $\theta_y$ Detection, i.e. represent $\tau$ as illustrated in figure 11.

Practically, the rotation angles around 0 degrees plays more important role in our end product evaluation than that of around 90 degrees. Since Bookstein constraint and Gauss-Newton method shows more scatterings around 0, we select Trace constraint based ellipse fitting method as our candidate.

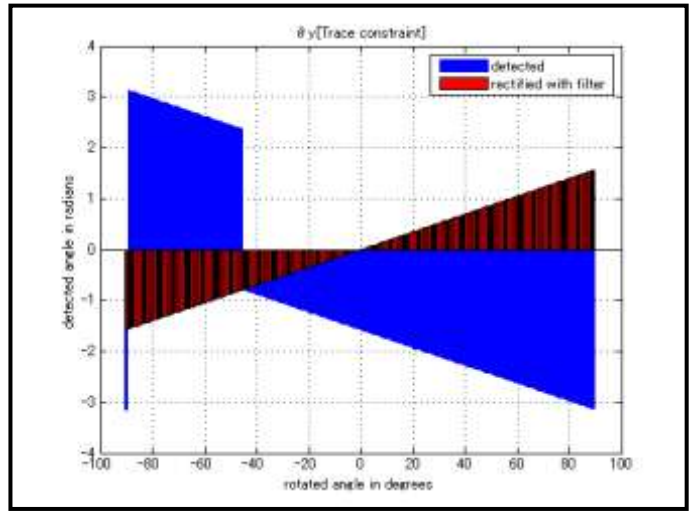

Figure 11. Rectified detection using filter

IV. SURFACE DIRECTION DETECTION

As proposed in figure 5, surface direction detection is proposed in this paper as a 2D line fitting problem of a slope estimation of the fitted line in ZY plane at specific evaluation point. The equation of the object line is presented in equation (5) bellow.

$$ax + by + c = 0 \quad (5)$$

Since intersection is zero, the surface direction angle will be,

$$\alpha = \pi/2 - \tan^{-1}(-a/b) \quad (6)$$

Fitting are performed to obtain values of 'a', 'b' and 'c' of above equation. We have two main options, using maximum likelihood estimator or least square estimator. We used the latter for our testing. In future after checking the behaviour of measurement errors, we experiment on the former as well. If the errors would satisfy following variance in equation (7), one could expect similar results from both estimator types.

$$\sigma_y^2 = \sigma_z^2 = \sigma^2 \quad (7)$$

Statistically, this is the use of total least squares method, i.e. the sum of the orthogonal squared distances from the data points to the fitted line should be in minimum. In other terms this is using orthogonal least square estimator. The following figure 12 illustrates the fitting results. We designed a simple filter to rectify the detection using the sign of 'a' from equation (5). The rectified detection data are illustrated in figure 13.

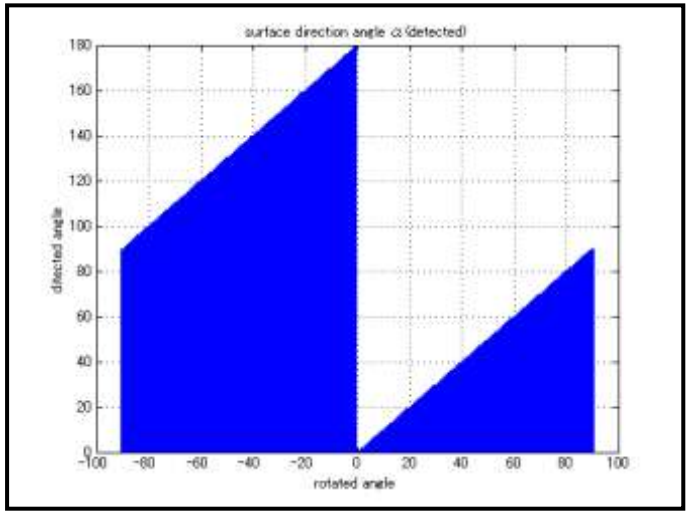

Figure 12. Fitting results for surface direction angle

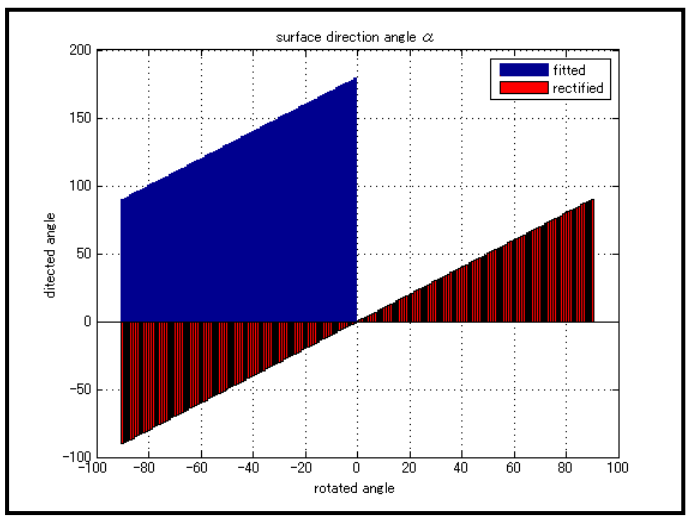

Figure 14. Rectified fitting results

## CONCLUSIONS

In this research we addressed a computation approach of an elliptical helical pipe bending problem in end product evaluation. Although the actual product is still under development, we can present that our approach is good enough to proceed with.

As future prospects for this product evaluation computation, it is worth trying detection at any arbitrary point in 3D. Also worth applying algorithms such as Genetic Algorithm to compare results. How ever in our computation we avoided possible computation cost by utilizing linear methods as possible. The usual case is unpredictable nonlinearities do occur in real world machining. Therefore precision accuracy could be addressed only after real machining.

## ACKNOWLEDGMENT

The authors express their sincere gratitude to Kikuchi Seisakusho Co. Ltd. at Hachioji-City, Tokyo for the financial and technical support for this project.